\documentclass[aps]{revtex4}
\usepackage{graphicx}
\usepackage[all]{xy}
\usepackage{amsmath}
\usepackage{amssymb}
\usepackage{color}
\usepackage[latin1]{inputenc}

\newcommand{\bes}{\begin{subequations}}
\newcommand{\ees}{\end{subequations}}
\newcommand{\benn}{\begin{eqnarray*}}
\newcommand{\eenn}{\end{eqnarray*}}

\def\ben{\begin{eqnarray}}
\def\een{\end{eqnarray}}
\def\be{\begin{equation}}
\def\ee{\end{equation}}

\begin{document}
\title{Localized fields on scalar global defects}
\author{F.A. Brito$^{a}$ and { H.H.B. Silva}$^{a,b}$}
\affiliation{{\small {{$^a$}{Departamento de F\'\i sica, Universidade Federal de Campina
Grande, \\Caixa Postal 10071, 58109-970 Campina Grande, Para\'\i ba,
Brazil}}\\
{$^b$}{Departamento de F\'\i sica Te\'orica e Experimental, Universidade Federal do Rio
Grande do Norte, 59072-970 Natal, Rio Grande do Norte,
Brazil}}}

%\date{\today}

\begin{abstract}
We investigate the localization of modes on the worldvolume of a $p$-brane embedded in $p+d+1$-dimensional 
spacetime. The $p$-brane here is such that its profile is regarded as a scalar global defect and
the localized modes are scalar modes that come from the fluctuations around such defect. The effective action
on the brane is computed and the induced potentials are typically $\phi^4$-type potentials that are flatter for
lower $d$-dimensions. We also make a connection of such scalar global defects with black $p$-branes in certain
limits.
\end{abstract}

%\pacs{11.27.+d}

\maketitle

\section{Introduction} % {12-1, 13, 14, 14-1, 14-2}

{ In the present investigation we are interested in a mechanism of compactifying higher dimensional spacetime down to lower dimensional one.  There are already in the literature several results in high energy physics whose interest is the understanding why our Universe presents only three spatial dimensions. This was specially considered in Refs.~\cite{12-1,13,14} where the authors considered diluting modes in string gas \cite{12-1} and D-brane gas \cite{13, 14} cosmology to point out that these configurations may favor three-dimensional Universe. In the study \cite{bbl2006epl} one of the present authors also reported new ideas on this problem. In Ref.~\cite{bbl2006epl} it was basically considered the aforementioned idea of space-time being filled with a D-brane gas \cite{13,14}. For more recent investigations along these lines see \cite{Herbut:2012ep,Altshuler:2012ss,Bhowmick:2012dc,Clifton:2011jh,Eingorn:2011vu,Bhowmick:2010dd,Ferrer:2009pq,Carroll:2009dn,Flachi:2009uq,Nelson:2008uy,Bhowmick:2008cq,Nelson:2008sv}. It was supposed that the brane gas dilutes into a nine-dimensional system, regarded as an oscillatory system with nine non-compact dimensions. The branes were manifested as phase fronts separating different phase states of oscillations \cite{14-1,14-2}. Such a system usually allows for bifurcations to form brane junctions considered to be a natural way to select lower dimensions. The key point presented was the fact that there is the possibility of the ten-dimensional spacetime to be filled with a network of junctions of higher dimensional co-dimension one branes intersecting orthogonally to form lower dimensional objects. Thus, junctions of several spatial dimensions could occur to form $p$-brane junctions, such a way gravity, matter and gauge fields tend to live at these junctions with smaller dimensions. In other words, although superstrings predict the Universe has 10 dimensions, such spacetime can be filled with brane junctions that allows for the phenomenon of gravity, matter and gauge fields localization on some preferable spacetime with lower dimension. 

Let us now present a new route to achieve a space selection mechanism based on a unique $p$-brane modeled by a special system rather than considering intersections of them. However, as in the junction case, the cosmological issues here also develop special appeal. For instance, we shall investigate at which number of extra dimensions of the spacetime the induced {\it inflaton} potential on the $p$-brane is flatter to favor sufficient inflation in our Universe modeled by a 3-brane. As we shall discuss below this can be done through the use of scalar global defects \cite{Bazeia:2003qt}.

%As we discuss below, such space-time can be the 3 + 1 dimen- sional, as world-volume of 3-brane junctions that fills the ten dimensional space-time, in accord with Refs. [2, 3]. The space selection via brane junctions is endowed with the phenomenon of gravity, matter and gauge fields lo- calization on branes [6, 7] and brane junctions [8, 9].
}

Thus, in this work we investigate a compactification mechanism through localization of fields modes on $p$-brane embedded in $p+d+1$-dimensional 
spacetime with topology ${\cal M}^{p+d+1}=\mathbb{M}^{p+1} \times \mathbb{R}^d$. The $p$-brane here is such that its profile is regarded as a scalar global defect and
the localized modes are scalar modes that come from the fluctuations around the scalar global defects first introduced in \cite{Bazeia:2003qt}. These modes are described via eigenfunctions that satisfy a Schroedinger-like equation for the fluctuations. We then integrate out these modes
in the internal space $\mathbb{R}^d$ to obtain an effective action describing the fields localized on the scalar global defect (the $p$-brane) following the lines of the references \cite{Z2000,B2005,B2008,B2009}. Such
defects are scalar soliton solutions of a scalar field theory in arbitrary dimensions even for $d>2$. This is known to evade the Derrick's theorem with the price of being a theory that breaks translational invariance. They are different of the co-dimensional one objects mostly used in the braneworlds scenarios \cite{1,2,3,4,5,6,7,8,9,10,11,12, 12-a,12-b,12-c,12-d,12-e,12-ee,12-f,12-g}. The scalar potential in the Lagrangian now depends explicitly on the spatial coordinates. However, we show that the effective theory describing localized scalar modes on the global defect worldvolume recovers the translational invariance because the effective potentials has no dependence on any spatial coordinate on the $p$-brane. In this sense we conclude that in this compactification process arises a mechanism in which a theory with broken translation invariance in higher dimensions recovers the translational invariance in lower dimensions.

{ This Letter is organized as follows. In Sec.~\ref{eff} we compute the effective action for scalar global defect spectrum. We find a general form for the effective potential localized on a $p$-brane. We also find a connection between the scalar global defects and black $p$-branes in certain limits. Finally, in Sec.~\ref{concu} we present our final conclusions. }

% {\color{red} This is in accord with the recent considerations on spacetime symmetries broken in high energy probing extra dimensional physics in our Universe.}

\section{Effective action for the scalar global defect spectrum}
\label{eff}

Consider a theory of a scalar field embedded in a $p+d+1$-dimensional spacetime with topology
\begin{equation} \label{C1}
{\cal M}^{p+d+1}=\mathbb{M}^{p+1} \times \mathbb{R}^d,
\end{equation}
where $M=(y^{\mu},q)$, with $q=(x_1,x_2,...,x_d)$ being coordinates in the internal flat space and $y^{\mu}=(t,y^i)$
$(i=1,2,...,p)$ are coordinates of the $p$-brane worldvolume embedded in a $(p+d+1)$-
dimensional spacetime. The action is given by 
\begin{equation}\label{C2}
S=\int
d^{p+1}yd^dq\left[\frac{1}{2}\partial_{M}\phi\partial^{M}\phi-V(\phi)\right],
\end{equation}
that can be written in a more convenient form as follows
\begin{equation}\label{C3}
S=\int
d^{p+1}yd^dq\left\{\frac{1}{2}\left[\left(\frac{\partial\phi}{\partial
t}\right)^2-\left(\sum_{i=1}^{p}\frac{\partial\phi}{\partial
y_i}\right)^2-\left(\nabla_q\phi\right)^2\right]-V(\phi)\right\},
\end{equation}
with $d^dq=dx_1dx_2...dx_d=r^{d-1}drd\Omega_{(d-1)}$, being
$\Omega_{(d-1)}=\frac{2\pi^{d/2}}{\Gamma(d/2)}$ the $(d-1)$-dimensional volume, of the unit $(d-1)$-sphere.

Let us now apply the perturbation theory to the scalar field $\phi$ as
\begin{equation} \label{C4}
\phi(q,y^{\mu}) \longrightarrow \bar\phi(q)+\eta(q,y^{\mu}),
\end{equation}
such that
\begin{equation} \label{C5}
S(\phi) \longrightarrow S(\bar\phi,\eta).
\end{equation}
This allows us to expand the potential around the static solution describing the scalar global defect.
The action now reads 
\begin{eqnarray} \label{C6}
S&=&\int
d^{p+1}yd^dq\left\{-\frac{1}{2}\left(\nabla_q\bar\phi\right)^2-V(\bar\phi)-
\frac{1}{2}\partial_\mu\eta\partial^{\mu}\eta \right.
\nonumber\\
&+&\left.\frac{1}{2}\left(\eta\nabla_q^{2}\eta-\eta
V''(\bar\phi)\eta\right)-
\frac{V'''(\bar\phi)}{3!}\eta^3-\frac{V''''(\bar\phi)}{4!}\eta^4+...\right\},
\end{eqnarray}
that is an action for the fluctuations $\eta$ of the $p$-brane. Since we assume the scalar field and fluctuations with spherical symmetry, the Laplacian is simply given in terms of the radial coordinate
\begin{equation}
\nabla^2_q=\frac{1}{r^{d-1}}\frac{d}{dr}\left(r^{d-1}\frac{d}{dr}\right).
\end{equation}
Now, by using this action, we can identify two important terms: the tension $T_p$ of
the $p$-brane and the bilinear Hamiltonian operator $\eta H\eta$, i.e.,
\begin{equation} \label{C7}
T_p=\int
d^dq\left[\frac{1}{2}(\nabla_q\bar\phi)^2+V(\bar\phi)\right]
\end{equation}
and
\begin{equation} \label{C8}
\frac{1}{2}\eta
H\eta=\frac{1}{2}\eta(-\nabla_q^2+V''(\bar\phi))\eta.
\end{equation}
Thus,
\begin{eqnarray} \label{C9}
S&=&-\int d^{p+1}yT_p+\int
d^{p+1}yd^dq\left[-\frac{1}{2}\partial_\mu\eta\partial^{\mu}\eta-\frac{1}{2}\eta
H\eta \right.
\nonumber\\
&-&\left.\frac{V'''(\bar\phi)}{3!}\eta^3-\frac{V''''(\bar\phi)}{4!}\eta^4+...
\right].
\end{eqnarray}
This action allows us to write a Lagrangian for the fluctuations, given by 
\begin{eqnarray} \label{C10}
{\cal L}_{(p+d+1)}&=&T_p
\delta^d(q)+\left[-\frac{1}{2}\partial_\mu\eta\partial^{\mu}\eta-\frac{1}{2}\eta
H\eta \right.
\nonumber\\
&-&\left.\frac{V'''(\bar\phi)}{3!}\eta^3-\frac{V''''(\bar\phi)}{4!}\eta^4+...
\right].
\end{eqnarray}
Using the Euler-Lagrange equation
\begin{equation}
\frac{\partial {\cal
L}_{(p+d+1)}}{\partial\eta}-\partial_{\mu}\left[\frac{\partial {\cal
L}_{(p+d+1)}}{\partial(\partial_{\mu}\eta)}\right]=0
\end{equation}
we find the following equation of motion
\begin{equation} \label{C11}
H\eta+\frac{V'''(\bar\phi)}{2!}\eta^2+\frac{V''''(\bar\phi)}{3!}\eta^3+...=\partial_{\mu}\partial^{\mu}\eta\equiv
\Box_{(p+d)}\eta.
\end{equation}
In the linear regime we have 
\begin{equation} \label{C12}
H\eta=\Box_{(p+d)}\eta.
\end{equation}

Now writing the fluctuations in terms of a sum of normal modes we find
\begin{equation} \label{C13}
\eta(y^{\mu},q)=\sum_{n}\xi_n(y)\psi_n(q),
\end{equation}
that substituting into (\ref{C12}) and assuming that the modes $\xi_n(y)$ are
fields describing localized particles on the $p$-brane satisfying a Klein-Gordon equation 
\begin{equation} \label{C14}
\Box_{(p+d)}\xi_n(y)=M_n^2\xi_n(y),
\end{equation}
we find a Schroedinger-like equation that governs the masses $M_n^2$ of the particles given by
\begin{equation} \label{C15}
-\nabla_q^2\psi_n(q)+V''(\bar\phi)\psi_n(q)=M_n^2\psi_n(q).
\end{equation}
On the other hand, assuming the following orthogonality condition for the wave functions $\psi(q)$
\begin{equation} \label{C16}
\int d^dq \psi_m(q) \psi_n(q)=\delta_{m,n},
\end{equation}
substituting (\ref{C13}) into the action (\ref{C9}) and finally integrating in $q$, we find the effective action 
\begin{equation} \label{C17}
S=-\int d^{p+1}y
\left[T_p+\sum_{n=0}^{N}\partial_{\mu}\xi_n(y)\partial^{\mu}\xi_n(y)+V(\xi)\right],
\end{equation}
where the potential for the localized modes is written as
\begin{equation} \label{C18}
V(\xi)=\int d^dq\left[\frac{1}{2}\eta
H\eta+\frac{V'''(\bar\phi)}{3!}\eta^3+\frac{V''''(\bar\phi)}{4!}\eta^4+...
\right].
\end{equation}

We now focus on scalar global defects by deforming the usual scalar field theory by considering the scalar potential with an explicit dependence on the spatial coordinates \cite{Bazeia:2003qt}
\begin{equation}  \label{C24}
V(\phi,r)=\frac{1}{2r^{2d-2}}W_{\phi}^2.
\end{equation}
The first derivative of the superpotential ($W_{\phi}=\partial
W/\partial\phi$) is given by
\begin{equation} \label{C25}
W_{\phi}=\left(\phi^{\frac{a-1}{a}}-\phi^{\frac{a+1}{a}}\right).
\end{equation}
$d>2$ is the dimension of the internal space and $a = 1,3,5...$ is a dimensionless parameter of the theory. To find topological solutions 
that describe scalar global defects where we shall introduce fluctuations around, we shall make use of the Bogomol'nyi formalism which is
useful to reduce equations of motion to first order differential equations that are easier to integrate. Thus, for static fields in $d$ 
dimensions we find
\begin{equation} \label{C26}
\frac{d\phi}{dr}=\pm\frac{1}{r^{d-1}}W_{\phi}.
\end{equation}
By substituting (\ref{C25}) into equation above, for the plus sign, we find the solution
\begin{equation} \label{C27}
\bar\phi(r)=\tanh^a\left[\frac{1}{a}\left(\frac{r^{2-d}}{d-2}\right)\right].
\end{equation}

Now before address the issue of localized spectrum on the $p$-brane, we shall first show that in the present system there is only one bound state, a zero mode given by the eigenfunction $\psi_0$, followed by a tower of continuum massive modes that we disregard in the present study.  It is not difficult to show that the Schroedinger-like equation can be factored in terms of another operator as follows \cite{Bazeia:2003qt}
\begin{equation} \label{C27a}
H=\frac{1}{r^{2d-2}}Q^{\dag} Q
\end{equation}
with
\begin{equation} \label{C27b}
Q=r^{d-1}\frac{d}{dr}\mp W_{\phi\phi},
\end{equation}
and
\begin{equation} \label{C27c}
Q^{\dag}=-r^{d-1}\frac{d}{dr}\mp W_{\phi\phi}.
\end{equation}
This guarantees that the operator $H$ is quadratic and so no tachyonic modes are allowed. Furthermore since the Schroedinger potential approaches zero as $r\to\infty$ then the only bound state is the zero mode $\psi_0$.
We can determine such a mode by solving the following eigenvalue equation
with $M_0=0$
$$H\psi_0=M_0^2\psi_0, \nonumber\\$$
that is,
$$\frac{1}{r^{d-1}}\left[r^{d-1}\frac{d}{dr}\mp W_{\phi\phi}\right]\psi_0=0,$$
whose solution is
\begin{equation} \label{C28}
\psi_0=c\exp\left[\pm\int\frac{W_{\phi\phi}}{r^{d-1}}dr\right],
\end{equation}
where $c$ is a normalization constant and in general just one solution with a particular signal $\pm$ is normalizable.
Thus, since we have only one bound state $\psi_0$, the effective action (\ref{C17}) reads
\begin{equation} \label{C29}
S=-\int d^{p+1}y
\left[T_p+\partial_{\mu}\xi_0(y)\partial^{\mu}\xi_0(y)+V(\xi)\right],
\end{equation}
the tension (\ref{C7}) is now written as
\begin{eqnarray}\label{C30}
T_p&=&\int
d^dq\left[\frac{1}{2}(\nabla_q\bar\phi)^2+V(\bar\phi)\right]=\frac{2\pi^{d/2}}{\Gamma{(d/2)}}\frac{2a}{(4a^2-1)}\nonumber\\
&=&\Omega_{(d-1)}\frac{2a}{(4a^2-1)},
\end{eqnarray}
and the effective potential (\ref{C18}) in now given by
\begin{equation} \label{C31}
V(\xi_0)=\int
d^dq\left[\frac{V'''(\bar\phi)}{3!}(\xi_0\psi_0)^3+\frac{V''''(\bar\phi)}{4!}(\xi_0\psi_0)^4+...\right].
\end{equation}
Notice that since we have just a zero mode, the mass term in the potential disappeared. The dots mean higher order terms. These terms are not 
present for $a=1$ since in this case $V(\phi)$ is at most of the fourth order, so that this potential turns out to be exact. However, the wave function is normalized only for $a>1$. 

\subsection{The connection with  black $p$-branes}

In spite of this, we use the case $a=1$ to show that our setup has a hidden connection with black $p$-branes as a solution of the type II supergravity. We do not need fluctuations (the wave function) for the moment. For this proposal let us compare our action (\ref{C2}) with  the bosonic sector of type II supergravity action (with $p+1+d=10$)
\begin{equation}\label{sugra}
S=\int
d^{p+1+d}x\sqrt{g}\left[e^{-2\Phi}(R+4\partial_{M}\Phi\partial^{M}\Phi)-\frac12|F_{p+2}|^2\right].
\end{equation}
The extremal black D$p$-brane solution is given by 
\begin{equation}\label{sugra-sols}
ds^2=H_p^{-1/2}\eta_{\mu\nu}dy^\mu dy^\nu + H_p^{1/2}(dr^2+r^2d\Omega_{8-p}^2),\qquad H_p(r)=1+\left(\frac{r_p}{r}\right)^{7-p}, \qquad 
e^{\Phi}=g_s H_p^{(3-p)/4},
\end{equation}
with the R-R field strength given by $F_{p+2}=dH^{-1}_p\wedge dx^0\wedge dx^1\wedge ...\wedge dx^p$. In the limit $r\gg r_p$, $H_p\to1$,
the dilaton $\Phi$ approaches to a constant and $|F_{p+2}|\sim {1}/{r^{8-p}}$. Now substituting this solution into the action (\ref{sugra}) we find
\begin{equation}\label{sugra-sols-lim}
S=-\frac12\Omega_{8-p}\int
d^{p+1}y\int\frac{dr}{r^{8-p}}.
\end{equation}
The same action can be found from (\ref{C2}) by considering $a=1$ such that $V=W_\phi^2/2r^{2d-2}=(1-\phi^2)^2/2r^{2d-2}$ with the solution (\ref{C27}) gives 
\begin{equation} \label{new-pot}
V=\frac{1}{2r^{2d-2}}{\rm sech}^4{\left[\frac{1}{a}\left(\frac{r^{2-d}}{d-2}\right)\right]}.
\end{equation}
For $r\to\infty$ we find $V\to{1}/{2r^{2d-2}}$ and $\bar{\phi}\sim 1/r^{d-2}$, such that substituting this solution into (\ref{C2})
we find 
\begin{equation}\label{soliton-sols-lim}
S=-\frac12\Omega_{d-1}\int
d^{p+1}y\int\frac{dr}{r^{d-1}}.
\end{equation}
Of course, the actions (\ref{sugra-sols-lim}) and (\ref{soliton-sols-lim}) are the same for $p+d+1=10$.
Notice that the heaviest $p$-branes, for fixed $a$, occur for $d\simeq 7$. Being $M=T_p V_p$, the entropy of the corresponding black $p$-brane $S\sim M^2$ is also maximal around this dimension.

\subsection{Realization of flatter effective potentials for lower internal dimensions}

Let us now study the cases with $a>1$. The effective potential ({\it inflaton potentials}) develops $\xi^3$-terms only for $a=2$, but in the following we discuss only cases for $\xi^4$-terms. We shall focus only on the cases $d=3$ and  $d=6$ with $a=3$ to compare flatness of the induced $\xi^4$-terms in the effective potential which is essential for cosmological issues. 

The effective action for $d=3$ and $a=3$ is found by considering the following results:
\begin{itemize}
\item {Static solution}
\end{itemize}
\begin{equation} \label{C40a}
\bar\phi=\tanh^3\left(\frac{1}{3r}\right)
\end{equation}
\begin{itemize}
\item {Normalized wave-function}
\end{itemize}
\begin{equation} \label{C41}
\Psi_0=2.057\tanh^2\left(\frac{1}{3r}\right){\rm
sech}^2\left(\frac{1}{3r}\right)
\end{equation}
\begin{itemize}
\item {$p$-brane tension}
\end{itemize}
\begin{equation} \label{C42}
T_p=\frac{24}{35}\pi
\end{equation}
\begin{itemize}
\item {Effective potential}
\end{itemize}
\begin{equation} \label{C43}
V(\xi_0)=0.95\pi\xi_0^4.
\end{equation}

Similarly, the effective action for $d=6$ and $a=3$ is found by considering the following results:
\begin{itemize}
\item {Static solution}
\end{itemize}
\begin{equation} \label{C43a}
\bar\phi=\tanh^3 \left(\frac{1}{12r^4}\right)
\end{equation}
\begin{itemize}
\item {Normalized wave-function}
\end{itemize}
\begin{equation} \label{C44}
\Psi_0=6.7009\tanh^2\left(\frac{1}{12r^4}\right){\rm
sech}^2\left(\frac{1}{12r^4}\right)
\end{equation}
\begin{itemize}
\item {$p$-brane tension}
\end{itemize}
\begin{equation} \label{C45}
T_p=\frac{6}{35}\pi^3
\end{equation}
\begin{itemize}
\item {Effective potential}
\end{itemize}
\begin{equation} \label{C46}
V(\xi_0)=26.72\pi^3\xi_0^4.
\end{equation}

\begin{figure}[h!]
\centerline{\includegraphics[{angle=90,height=6.5cm,angle=270,width=6.5cm}]{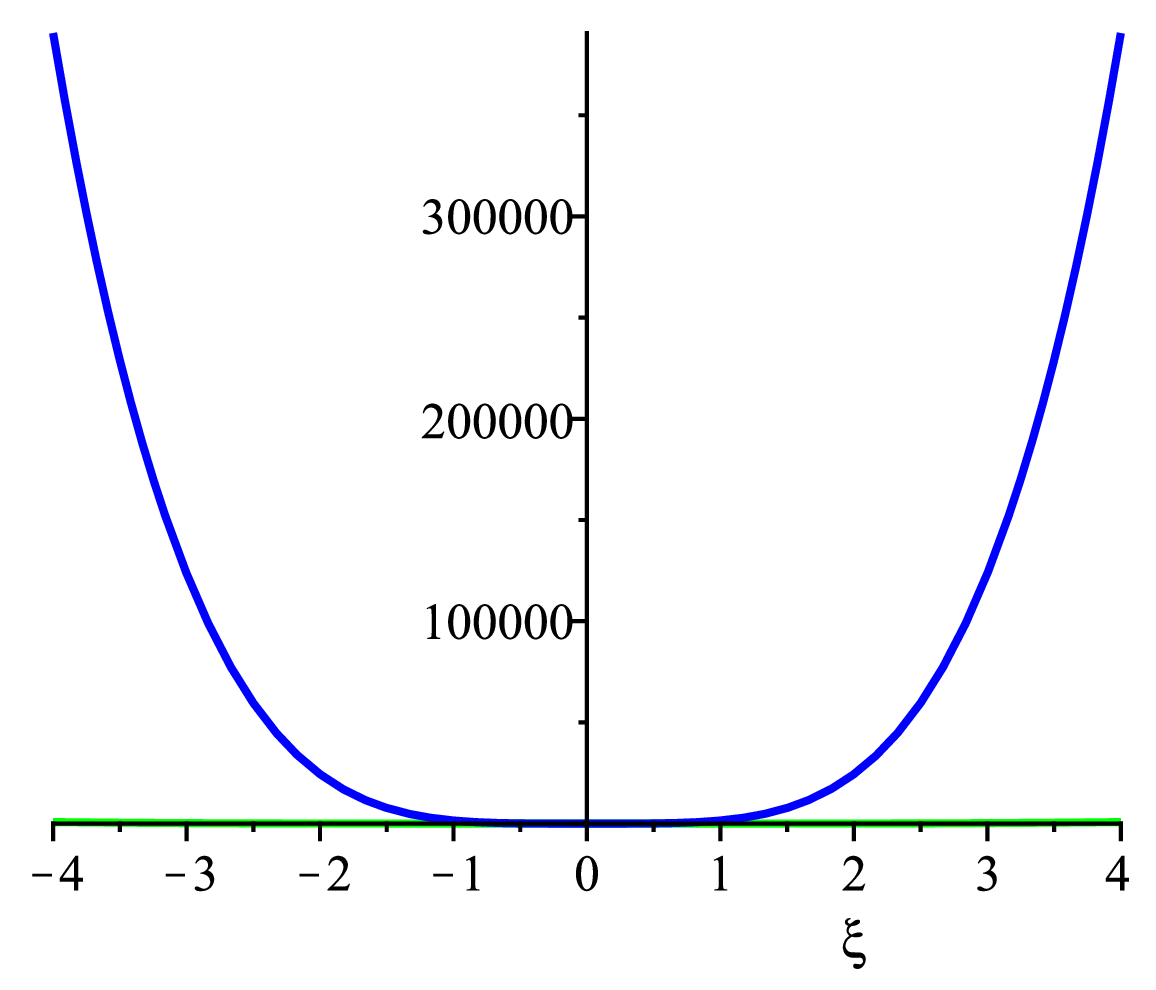}}
\caption{Effective potential for the mode $\xi_0$ in $d=3$ (green) is flatter than in $d=6$ (blue).}\label{fig-pot}
\end{figure}

The Fig.~\ref{fig-pot} shows how flat the effective potential on the $p$-brane is for dimensions $d\!=\!3$ and $d\!=\!6$.

\section{conclusions}
\label{concu} 

{In summary, our space selection mechanism works as follows. The effective potentials (the {\it inflaton potentials}) are such that they are flatter for lower $d$-dimension. 
This issue is important in inflationary scenarios since flatter potentials may produce sufficient inflation. 
We can estimate the suitable $d$-dimensional internal space to produce sufficient inflation in our Universe
($p=3$)-brane. For instance, in $d=3$ is easier to produce inflation than $d=6$, this may explain why we should expect a small number of 
large extra dimensions to develop. Thus, since inflation is crucial for the development our Universe, the more we have sufficient inflation the less
a number of extra dimensions is favored. We also conclude that although the scalar global defects break the 
translation symmetry because the scalar potential has explicit dependence with
the spatial coordinates, it is possible to find effective theories living on
the scalar global defect (the $p$-brane) worldvolume whose effective potentials are invariant
under such symmetry.  This is so because such potentials have no explicit dependence
with the coordinates of the $p$-brane, thus they may respect the translation symmetry along
the $p$-brane. In this sense we can realize that in this compactification process also appears 
a mechanism in which a higher dimensional theory that breaks translation symmetry may
recover this symmetry in lower dimensions. 
%This effect may also be used to probe extra dimensions in our Universe.
}
 
The authors would like to thank CAPES,
CNPq, and PROCAD/PNPD-CAPES for partial support.

%%%%%%%%%%%%%%%%%%%%%%%%%%%%%%%%%%%%%%%%%%%%%%%%%%%%%%%

\end{document}